\documentclass[aps, prd,
nofootinbib,
twocolumn,
superscriptaddress,
floatfix
]{revtex4-2}

\usepackage{amsmath}
\allowdisplaybreaks
\usepackage{graphicx}
\usepackage{longtable}
\usepackage{multirow}
\usepackage{epsfig} 
\usepackage[usenames]{color}
\usepackage[
colorlinks=true,
linkcolor=blue,
citecolor=blue,
urlcolor=blue
]{hyperref}
\usepackage{colortbl}
\usepackage{xcolor}
\definecolor{cream}{RGB}{253, 246, 227}
\pagecolor{cream!90}

\newcommand*{\dif}{\mathop{}\!\mathrm{d}}

\begin{document}
\title{Bulk viscosity of quark matter across the QCD phase transitions}

\author{Chong-long Xie}
\affiliation{ MOE Key Laboratory for Non-equilibrium Synthesis and Modulation of Condensed Matter,
	School of Physics, Xi’an Jiaotong University, Xi’an 710049, China}
\author{Guo-yun Shao}   
\email[ ]{gyshao@mail.xjtu.edu.cn} 
\affiliation{ MOE Key Laboratory for Non-equilibrium Synthesis and Modulation of Condensed Matter, 
	School of Physics, Xi’an Jiaotong University, Xi’an 710049, China}
\author{Ming-zheng-xuan Wu}
\affiliation{ MOE Key Laboratory for Non-equilibrium Synthesis and Modulation of Condensed Matter,
	School of Physics, Xi’an Jiaotong University, Xi’an 710049, China}
\author{Wei-bo He}
\affiliation{ MOE Key Laboratory for Non-equilibrium Synthesis and Modulation of Condensed Matter,
	School of Physics, Xi’an Jiaotong University, Xi’an 710049, China}

\begin{abstract}
Based on the kinetic theory with relaxation time approximation, we investigate the bulk viscosity ($\zeta$) and its ratio to shear viscosity  ($\zeta/\eta$) of quark matter at finite temperature and chemical potential with the in-medium particle masses derived in the 2+1 flavor Polyakov-loop improved Nambu--Jona-Lasinio (PNJL) model. We explore the behaviors of specific bulk viscosity ($\zeta/s$) and $\zeta/\eta$ across different QCD phase transitions, including the Mott phase transition, the chiral crossover, and the first-order transition with the associated metastable phase.  The calculation shows that both $\zeta/s$ and $\zeta/\eta$ are extremely small at high temperatures, approaching the nature of a conformal theory. Larger $\zeta/s$ and $\zeta/\eta$ are derived near the chiral phase transition at finite temperature. Along the chiral crossover line, $\zeta/s$ and $\zeta/\eta$ generally increase with decreasing temperature, though $\zeta/\eta$ exhibits a slight decline near the critical  endpoint (CEP). On the boundary of the first-order transition, $\zeta/s$ shows a non-monotonic variation with temperature. Furthermore, an additional peak structure emerges beyond the chiral phase boundary for both  $\zeta/s$ and $\zeta/\eta$, with magnitudes even exceeding those near the chiral crossover of $u, d$ quarks. Our analysis indicates this peak originates from the chiral crossover transformation of strange quark.
\end{abstract}

\maketitle
\section{Introduction}
\label{introduction}

The quark-gluon plasma (QGP) can be created in high-energy heavy-ion collision experiments.~Studying the transport properties of QGP provides important insights into the microscopic interactions governed by Quantum Chromodynamics (QCD). In hydrodynamics, the shear and bulk viscosities characterize the fluid's dissipative response to velocity gradients, measuring the departure from ideal local equilibrium behavior. Combined with high-energy experimental data from RHIC and LHC, particularly observables such as collective flow, higher-order flow harmonics, and jet quenching, hydrodynamic simulations reveal that the specific shear viscosity $\eta/s$ is remarkably small in high-energy collisions~\cite{romatschke2007viscosity, song2008suppression, denicol2009effect,heinz2013collective, gale2013hydrodynamic,parkkila2021bayesian,2019Bernhard,everett2021multisystem}, approaching the KSS lower bound of $1/4\pi$ derived in anti–de Sitter space/conformal
field theory (AdS/CFT)~\cite{kovtun2005viscosity, policastro2001shear}. This indicates that the QGP produced at high energy behaves as a nearly perfect fluid with strongly coupled characteristics~\cite{teaney2003effect, shuryak2004toward,shuryak2005rhic}.

The determination of QGP viscosity coefficients remains challenging issues in high-energy nuclear physics. 
The dependence of transport coefficients on temperature and baryon chemical potential  reflects the non-perturbative characteristics of QCD, and is related to the QCD  phase transitions~\cite{sasaki2010transport}. It has been conjectured that $\eta/s$ exhibits a minimum near the deconfinement  transition from QGP to hadrons, analogous to the behavior observed in substances such as water, helium, and nitrogen~\cite{csernai2006strongly,lacey2007has}. Different from the shear viscosity describing a ability to resist shape deformation,
the bulk viscosity coefficient $\zeta$  characterizes the energy dissipation during volume expansion.  
At extremely high temperature bulk viscosity is nearly negligible with $\zeta\ll\eta$~\cite{karsch2008universal}. As the temperature decreases near the phase transition, the enhancement of interactions and the change in degrees of freedom possibly lead to an increase in the bulk viscosity~\cite{2019Bernhard,Denicol09,karsch2008universal}. 

Numerous studies have calculated the viscous coefficients of quark matter with different methods (e.g.,~\cite{astrakhantsev2017temperature,haas2014gluon, christiansen2015transport, dubla2018towards,ghiglieri2018qcd,danhoni2023hot, gao2018temperature,zhang2018shear,chen2010shear,chen2013shear,sasaki2010transport,marty2013transport,soloveva2021shear,deb2016estimating,reichert2021first,deng2023shear,mclaughlin2022building,rehberg1996elastic, xiao2014bulk,mitra2017transport,moreau2019exploring,soloveva2020transport,mykhaylova2019quark,mykhaylova2021impact,shen2023viscosities,moore2008bulk,yang2023bayesian,abhishek2018transport,soloveva2022transport,singha2018calculations,madni2024shear,harutyunyan2017transport,Gardim20,Rojas24,Li2014dsa}). Considerable attention has been devoted to understanding the temperature dependence at small  chemical potentials. 
Compared to the shear viscosity coefficient, research on the bulk viscosity coefficient is relatively scarce~\cite{moore2008bulk,marty2013transport,chen2013shear,deb2016estimating,mitra2017transport,gao2018temperature,abhishek2018transport,singha2018calculations,soloveva2020transport,mykhaylova2021impact,Fu2013qra}, especially regarding the first-order phase transitions~\cite{sasaki2010transport,soloveva2022transport,Grefa,madni2024shear}.
The behavior of $\zeta$ on the chiral phase boundary has been calculated in the two-flavor NJL quark model~\cite{sasaki2010transport}, whereas the bulk viscosity on the phase boundary of the first-order transition associated with a spinodal structure is still lacking for three flavor quark matter.
In experiments, the Beam Energy Scan phase II program (BES-II) at RHIC-STAR  can probe a wide area of the QCD phase diagram, possibly covering  the QCD critical region and the first-order phase transition with the associated metastable phases~\cite{luo2017search,STAR2025zdq,STAR2025owm}. Therefore, the indepth research on viscosity coefficients of quark matter with QCD phase transition is necessary for constructing a complete hydrodynamic description of hot and dense QGP.

In Ref.~\cite{2024He}, we have detailedly investigated the shear viscosity across the QCD phase diagram and analyzed its relationship with QCD phase transitions. This work aims to study the dependence of specific bulk viscosity and the ratio $\zeta/\eta$ of quark matter on temperature, baryon density, and chemical potential, particularly their behavior near the QCD phase transitions, including the pion Mott phase transition, the chiral crossover, the critical region, and the first-order phase transition with a spinodal structure. The calculations will contribute to understanding the bulk viscous properties of QCD matter under different thermal parameters. It will also provide valuable reference  for hydrodynamic simulations of data from RHIC-BES II, 
aiding in the analysis of QCD phase transition. 

The calculations are conducted within the framework of kinetic theory under the relaxation time approximation.  
We employ the 2+1 flavor PNJL model to calculate the in-medium quasiparticle masses, with which the scattering cross sections under different thermal parameters are derived. 
The paper is organized as follows. In Sec.~II, we introduce the framework to calculate the bulk viscosity in kinetic theory with the relaxation time approximation. In Sec.~III, we illustrate the numerical results of specific bulk viscosity ($\zeta/s$) and the ratio  $\zeta/\eta$ under different physical conditions, and discuss the relationship with QCD phase transitions. A summary is given in Sec. IV.

\section{Framework}

Within the framework of relativistic kinetic theory, the transport coefficients characterize the first-order gradient expansion of the off-equilibrium spatial components of the energy-momentum tensor. The bulk and shear viscous coefficients, $\zeta$ and $\eta$,  can be derived by equating the microscopic definition of the stress tensor with its macroscopic counterpart in viscous hydrodynamics.
Following the detailed derivation within  the relaxation time approximation\cite{albright2016quasiparticle,mykhaylova2021impact}, the bulk viscous coefficient for quark matter can be described with
\begin{align}
	\label{eq_bulk}
	\zeta ={}&\frac{1}{9T}\sum_{i=q,\bar{q}}
	\tau_{i}d_{i} \int\frac{\dif^{3}p}{(2\pi)^3} \frac{1}{E_{i}^{2}}\notag\\
	{}&\left[{p}^{2}-3c_{n_B}^{2}T^{2}E_{i}\frac{\partial}{\partial T}\left(\frac{E_{i}-\mu_i}{T}\right)_{\sigma}\right]^{2}f^0_{i}(1-f^0_{i}).
\end{align}
In this formulation, $q(\bar{q}) = u, d, s (\bar{u}, \bar{d}, \bar{s})$ denotes  quarks and antiquarks; $\tau_i$ represents the relaxation time; $d_i = 2N_c$ is the degeneracy factor; 
and $E_{i} = \sqrt{p^2 + M_i^2}$ is the dispersion relation with a dynamical mass $M_i$ for (anti)quarks. The quark chemical potential $\mu_i = \mu_B/3$ is taken in the calculation.
The $c_{{n_B}}$ refers to the speed of sound at constant net baryon density $n_B~$\cite{2022He}, 
\begin{equation}
	c_{n_B}^2=\bigg(\frac{\partial P}{\partial\epsilon}\bigg)_{n_B}=\frac{s \chi_{\mu \mu}-n_B \chi_{\mu T}}{T\left(\chi_{T T} \chi_{\mu \mu}-\chi_{\mu T}^2\right)},
\end{equation}
where the second-order susceptibility $\chi_{x, y}$ is defined as $\chi_{x,y}=\partial^2 P / \partial x \partial y$. The $\sigma = s/n_B$ denotes the entropy per baryon, and $f^{0}_{i}$ stands for the equilibrium distribution function of (anti)quarks. The calcualtion will be conducted in the PNJL model with 
the distribution functions 
\begin{equation}\label{distribution}
	\!\!\!f^0_q\!=\!\frac{\!\Phi e^{\!-\!(\!E_i\!-\!\mu_i\!)\!/\!T}\!+\!2\bar{\Phi} e^{\!-\!2(\!E_i\!-\!\mu_i\!)\!/\!T}+e^{-3(E_i-\mu_i)/T}}
	{\!1\!+\!3\Phi e^{\!-\!(\!E_i\!-\!\mu_i\!)\!/\!T}\!+\!3\bar{\Phi} e^{\!-\!2(\!E_i\!-\!\mu_i)/T}\!+\!e^{\!-\!3(E_i\!-\!\mu_i)\!/\!T}}
\end{equation}
for quarks and
\begin{equation}\label{antif}
	\!\! f^0_{\bar q}\!=\!\frac{\! \bar \Phi e^{\!-\!(\!E_q\!+\!\mu_q\!)\!/\!T}\!+\!2\Phi e^{\!-\!2(\!E_q\!+\!\mu_q\!)\!/\!T}+e^{-3(E_q+\mu_q)/T}}
	{\!1\!+\!3\bar \Phi e^{\!-\!(\!E_q\!+\!\mu_q\!)\!/\!T}\!+\!3\Phi e^{\!-\!2(\!E_q\!+\!\mu_q\!)/T}\!+\!e^{\!-\!3(\!E_q\!+\!\mu_q)\!/\!T}}
\end{equation}
for antiquarks, where $\Phi$ and $\bar{\Phi}$ are the order parameters of the Polyakov loop and its conjugate, respectively.

The relaxation time $\tau_i$ plays a crucial role in determining the bulk viscosity, as it characterizes the timescale of microscopic scattering processes in the thermal medium. In this work, we consider the $2\to 2$ scatterings among $u$, $d$, and $s$ quarks and their antiparticles. 
The averaged relaxation time for particle species $i$ can be derived with
\begin{equation}
	\tau_i^{-1}\left(T,\mu_q\right)=\sum_{j=q, \bar{q}} n_j\left(T, \mu_q\right) \bar{w}_{ij},
\end{equation}
where $n_j\left(T, \mu_q\right)$ is the quark or antiquark number density of species $j$. $\bar{w}_{ij}$ is the  averaged transition rate defined as
\begin{equation}\label{wij}
\begin{aligned}
	\bar{w}_{i j}={}&\frac{1}{n_{i}n_{j}}\int\frac{\dif^{3}\mathbf{p}_i}{(2\pi)^3}
	\int\frac{\dif^{3}\mathbf{p}_j}{(2 \pi)^3}
	d_q f_i^{(0)}\left(E_i, T, \mu_q\right)d_q\\
	{}&\times f_j^{(0)}\left(E_j,T,\mu_q\right)\cdot v_{\mathrm{rel}}\cdot\sigma_{ij\to cd}\left(s,T,\mu_{q}\right).
\end{aligned}
\end{equation}
Here $v_{\mathrm{rel}}$ is the relative velocity and $\sqrt{s}$ is the center-of-mass energy. $\sigma_{ij\to cd}$ is the cross section from the initial incident (anti)quarks $(i, j)$ to  the final outgoing (anti)quarks $(c, d)$ at collision energy $\sqrt{s}$.

The relative velocity $v_{\mathrm{rel}}$  can be calculated in the c.m frame \cite{2012Eric,soloveva2021shear} with 
\begin{equation}
	v_{\mathrm{rel}}=\frac{\sqrt{\left(E_i^* E_j^*-\mathbf{p}_{i}^{*}\cdot\mathbf{p}_{j}^{*}\right)^2-\left(M_i M_j\right)^2}}{E_{i}^{*}E_{j}^{*}},
\end{equation}
where
\begin{equation}
	E_{i}^{*}=\frac{s+M_i^2-M_j^2}{2 \sqrt{s}},\,\,\,\,
	E_{j}^{*}=\frac{s-M_i^2+M_j^2}{2 \sqrt{s}} 
\end{equation}
and 
\begin{equation}
	\left|\mathbf{p}_i^*\right| =\frac{\sqrt{\left(s-\left(M_i+M_j\right)^2\right) \cdot\left(s-\left(M_i-M_j\right)^2\right)}}{2 \sqrt{s}} .
\end{equation}

The total cross section can be derived with the integral of the differential cross section
\begin{equation}\label{cs}
	\!\sigma\!=\!\int_{t_{-}}^{t^{+}}\!\!\! \mathrm{d} t \frac{\mathrm{d} \sigma}{\mathrm{d} t}\!\left(1\!-\!f^{(0)}(E_c^*,\! T, \mu_q)\!\right)\!\left(1\!-\!f^{(0)}(E_d^*,\! T\!, \mu_q)\!\right),
\end{equation}
where the factor $(1-f^{(0)})$ represents the Pauli blocking for quarks and/or antiquarks in the final state; $t$ is the Mandelstam variable; $t_{\pm}$ indicates the upper and lower limits of $t$ \cite{soloveva2021shear}
\begin{equation}
\begin{aligned}
	t_{ \pm}={}& M_i^2+M_c^2-\frac{1}{2 s}\left(s+M_i^2-M_j^2\right)\left(s+M_c^2-M_d^2\right)\\
	{}&\!\pm\!2\!\left.\sqrt{\!\frac{\left(s\!+\!M_i^2\!-\!M_j^2\right)^2}{4s}\!-\!M_i^2}\right.\! \sqrt{\frac{\left(s\!+\!M_c^2\!-\!M_d^2\right)^2}{4 s}\!-\!M_c^2} .
\end{aligned}
\end{equation}
The differential cross section in Eq.~(\ref{cs}) is written as 
\begin{equation}
	\frac{\dif\sigma}{\dif t}=\frac{1}{16\pi s_{ij}^{+}s_{ij}^{-}}\frac{1}{4N_{c}^{2}}\sum_{sc}|\mathcal{M}|^{2},
\end{equation}
where the  $s_{ij}^\pm$ is defined as
\begin{equation}
	s_{i j}^{ \pm}=s-\left(M_{i}\pm M_{j}\right)^2. 
\end{equation}
$ \frac{1}{4 N_c^2}\sum_{s c}|{\mathcal{M}}|^2$ denotes the matrix element squared averaged over the color and spin of the incident particles, and summed over the final scattered particles. The calculation of  $|{\mathcal{M}}|^2$ is related to the specific scattering channels, which can be performed using the standard methods in quantum field theory.  The medium-dependent quasiparticle masses of (anti)quarks as well as the exchanged mesons are calculated using the 2+1 flaovr PNJL model. A comprehensive description of this framework and calculation in the PNJL model are provided in our recent work~\cite{2024He}. One can also refer to Refs.~\cite{rehberg1996elastic,soloveva2021shear} for more details. 

\section{Numerical Results and Discussions}\label{results}
\subsection{Ratio of bulk viscosity to entropy density}

\begin{figure}[htbp]
	\centering
	\includegraphics[scale=0.4]{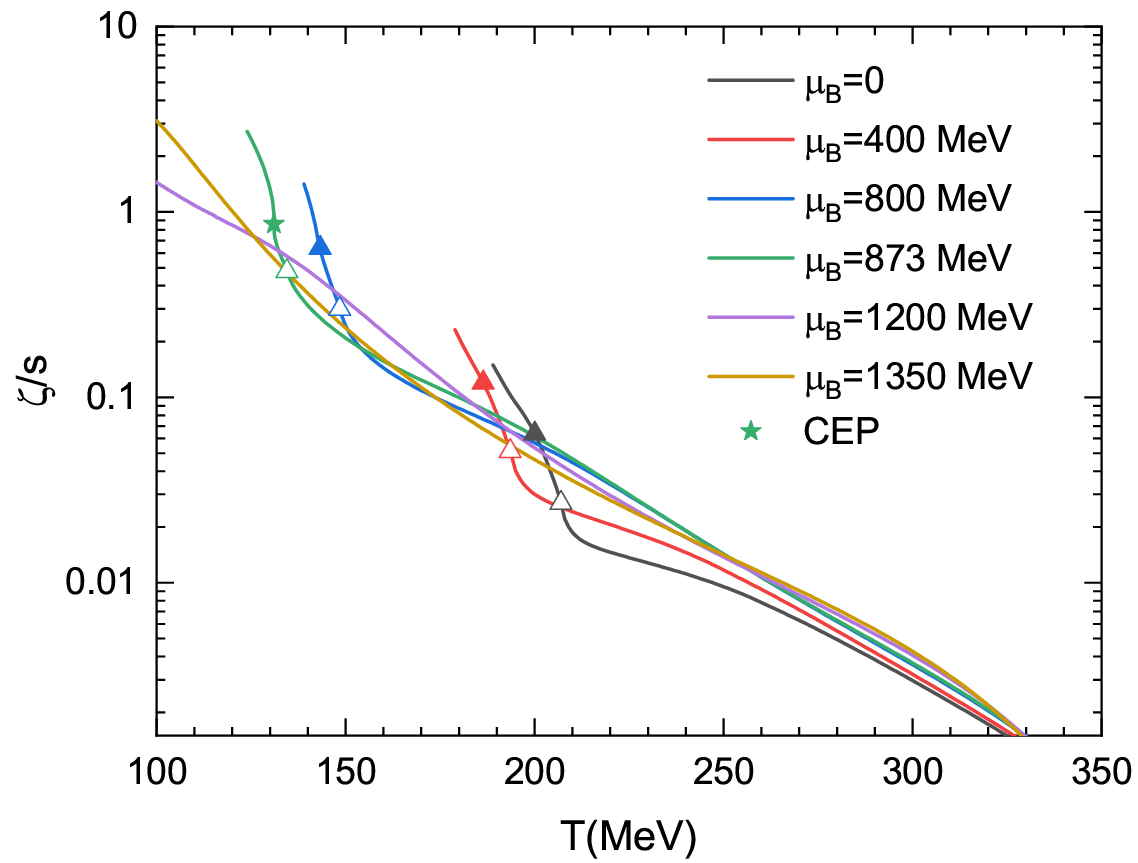}
	\caption{$\zeta/s$ as a function of temperature $T$ for baryon chemical potentials $\mu_{B} = 0$, 400, 800, 873, 1200, 1350 $\mathrm{MeV}$. The solid and hollow triangles represent  the $\zeta/s$ at the points where these $\mu_B(T)$ lines intersect the chiral crossover  line and pion Mott transition line, respectively.}
	\label{zetast}
\end{figure}

\begin{figure}[htbp]
	\centering
	\includegraphics[scale=0.4]{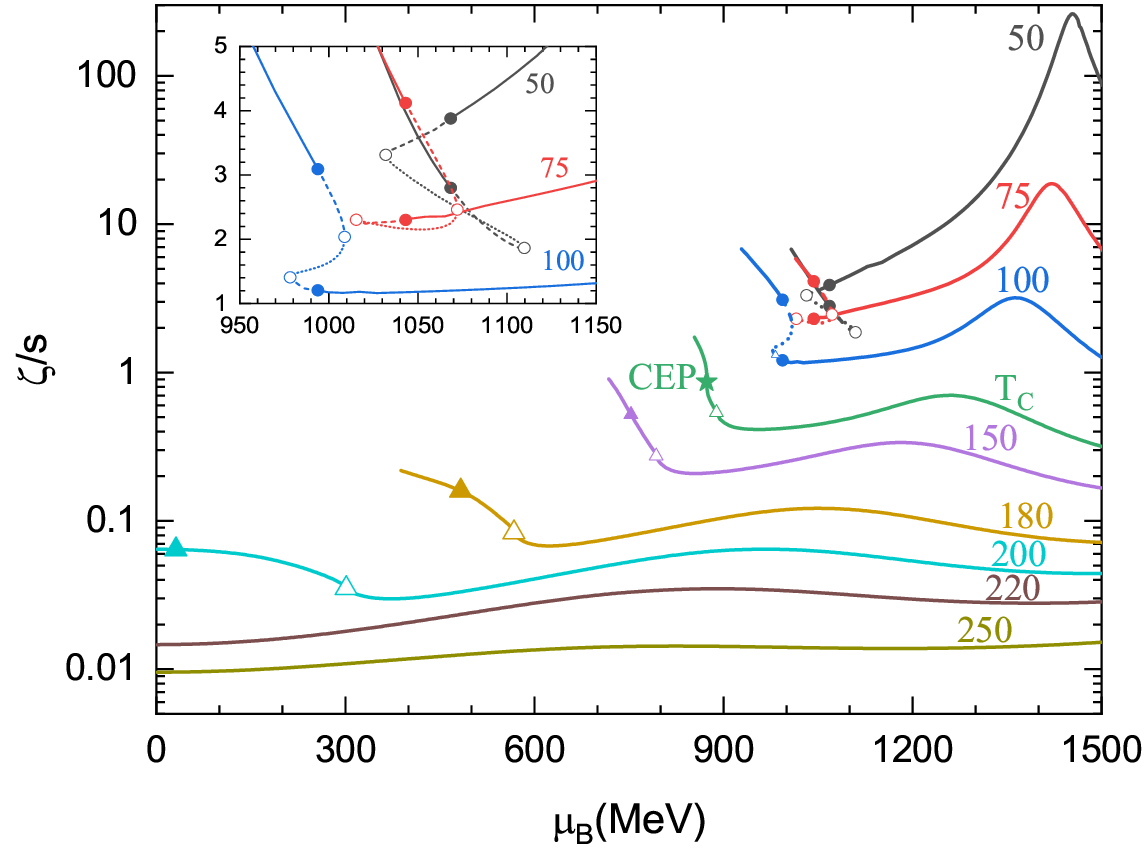}
	\caption{$\zeta/s$ as a function of $\mu_{B}$ for different temperatures. The solid and hollow triangles represent  the $\zeta/s$ at the points where these $T(\mu_B)$ lines intersect the chiral crossover line and pion Mott transition line, respectively.
	For $T=50, 75, 100\,$MeV, the two dots (cycles) on each isotherm correspond to respectively the high-density chirally restored phase and low-density chirally broken phase on the boundaries (spinodal line) of the first-order phase transition, and the dashed curves mark the metastable phases, including the high-density superheated phase and low-density supercooled phase.}
	\label{zetasmu}
\end{figure}

The specific bulk viscosity $\zeta/s$, which quantifies energy dissipation during the fireball expansion, serves as one of the essential inputs in hydrodynamic simulations of QGP evolution. The behavior of $\zeta/s$ of QGP  has a profound connection with the system's symmetry properties and equation of state characterized by the  speed of sound and interaction measurement.
The dependence of $\zeta/s$ on temperature and baryon chemical potential derived in the 2+1 flavor PNJL model are shown in Fig.~\ref{zetast} and Fig.~\ref{zetasmu}.

As observed in Fig.~\ref{zetast}, $\zeta/s$ exhibits an exponentially decrease with increasing temperature. It asymptotically approaches zero at extreme high temperatures, signaling the gradually restoration of scale invariance.
In a strict conformal field theory, scale invariance is necessarily preserved, and as a direct consequence, the specific bulk viscosity vanishes, i.e., $\zeta/s = 0$.  Whereas, the QCD containing particles with mass and complex interactions  is not a conformal field theory.  At high temperatures, the sources of conformal symmetry breaking, such as the trace anomaly and finite quark masses, become increasingly suppressed. As a result, the system approximates a scale-invariant one, and the specific bulk viscosity $\zeta/s$ approaches zero. This behavior serves as a  signature of the emergent near-conformal dynamics of QGP in the high-temperature limit. The similar result was derived in lattice QCD calculation~\cite{Astrakhantsev18}.

Fig.~\ref{zetast} also shows that $\zeta/s $ increase rapidly near the chiral crossover transformation~(marked by solid trangles) of $u,d$ quarks, approximately starting at the pion Mott dissociation (marked by blank trangles). This trend means the violation of approximate scale invariance with the increasing dynamical quark mass and larger interaction measurement (trace anomaly), $\theta=\epsilon-3P$, where $\epsilon$ is the energy density and $P$ is the pressure. 

In the calculation of bulk visicosity with the Kubo formula~\cite{kharzeev2008bulk} as well as the Eq.~(\ref{eq_bulk}), it shows that $\zeta$ is closely related to the speed of sound. For a scale-invariant relativistic system, the equation of state is uniquely determined by the relation $P = \epsilon / 3$. Consequently, the speed of sound is fixed at $c_s^2 = 1/3$, a value known as that of an ideal relativistic gas and a key thermodynamic signature of scale invariance.
While the bulk viscosity $\zeta$ quantifies the dynamical departure from scale invariance, the squared sound speed $c_s^2$ reflects the thermodynamic departure. The two measures are intrinsically linked, providing complementary insights into conformal symmetry breaking. Our previous calculation demonstrates that the squared speed of sound in QGP approaches the conformal limit of 1/3 at high temperatures and decreases significantly near the chiral crossover.

On the other hand, in the vicinity of the chiral crossover, the system's effective degrees of freedom change from quarks to hadrons, and the hadronic component prevails as temperature decreases. The hadronic model calculations predict that $\zeta/s$ rises with increasing temperature~\cite{noronha2009transport}. Consequently, combining insights from two different model frameworks indicates that the specific bulk viscosity coefficient likely reaches a maximum in the chiral crossover region~\cite{ryu2015importance}.

Fig.~\ref{zetasmu} depicts the dependence of $\zeta/s$ on baryon chemical potential $\mu_{B}$ for different temperatures. At high temperatures, $\zeta/s$ exhibits a mild  dependence on $\mu_{B}$.
With the decline of temperature, a rapid increase of $\zeta/s$ is observed near the chiral crossover region of $u,d$ quarks, with a starting point near the Mott dissociation of pion, as also shown in Fig.~\ref{zetast}. 
Besides, a distinct extremum appears at a larger baryon chemical potential on each isotherm. The numerical calculation identifies that this peak structure is driven by the chiral crossover of strange quark. 

\begin{figure}[htbp]
	\centering
	\includegraphics[scale=0.4]{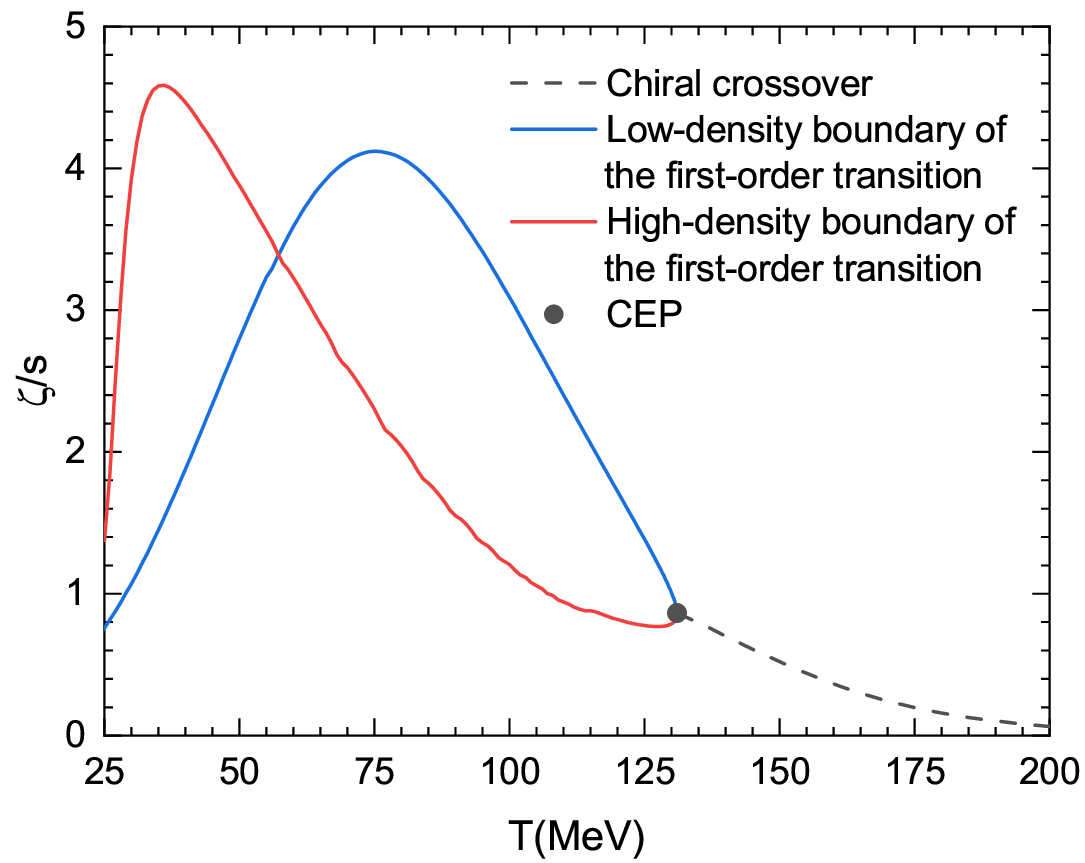}
	\caption{$\zeta/s$ on the chiral phase boundaries. }
	\label{zetaspt}
\end{figure}

These results further confirm that the specific bulk viscosity exhibits a stronger deviation from conformal invariance near the phase transition at finite temperature. Compared to the nearly negligible values at high temperatures, Fig.~\ref{zetast} and Fig.~\ref{zetasmu} show that $\zeta/s$ is significantly enhanced around the critical temperature, and continually increases as the temperature further decreases.
The bulk viscosity $\zeta$ essentially characterizes the efficiency with which the internal microscopic degrees of freedom redistribute energy and restore thermodynamic equilibrium under expansion.
A larger value of $\zeta/s$ indicates a slower relaxation process, meaning the system is less able to adjust to volume changes, leading to more  energy dissipation. 
Such slow relaxation is pronounced near the phase boundary and in the low-temperature regime, as given in our recent study~\cite{2024He}.

Fig.~\ref{zetasmu} also shows the behavior of $\zeta/s$ with a first-order phase transition for $T=100, 75, 50\,$MeV. 
It can be seen that, at $T=100$ and $75\,$MeV, the values of $\zeta/s $ on the phase boundary of the chirally restored high-density phase and the associated superheated metastable phase are smaller than those in the chirally broken low-density phase and the corresponding supercooled metastable phase. In contrast, the opposite trend is observed at $T=50\,$MeV. 
In general, the behavior of $\zeta/s $ across the first-order phase transition is affected  by temperature, baryon density (chemical potential), and dynamic quark masses.

Fig.~\ref{zetaspt} illustrates the behavior of $\zeta/s$ along the chiral phase boundary.
It can be seen that $\zeta/s$ increases monotonically along the chiral crossover transition line toward the CEP.
The two solid curves show $\zeta/s$ on the two boundaries of the first-order transition. The $\zeta/s$ on each branch exhibits a non-monotonic behavior. The numerical results indicate that both $\zeta$ and the entropy density $s$ on the boundary decline monotonically as the temperature decreases. The non-monotonicity in $\zeta/s$ arises from that the rates of decline for $\zeta$ and $s$ undergo an inflection with the lowering of temperature.
d{figure*}

\begin{figure}[htbp]
	\centering
	\includegraphics[scale=0.4]{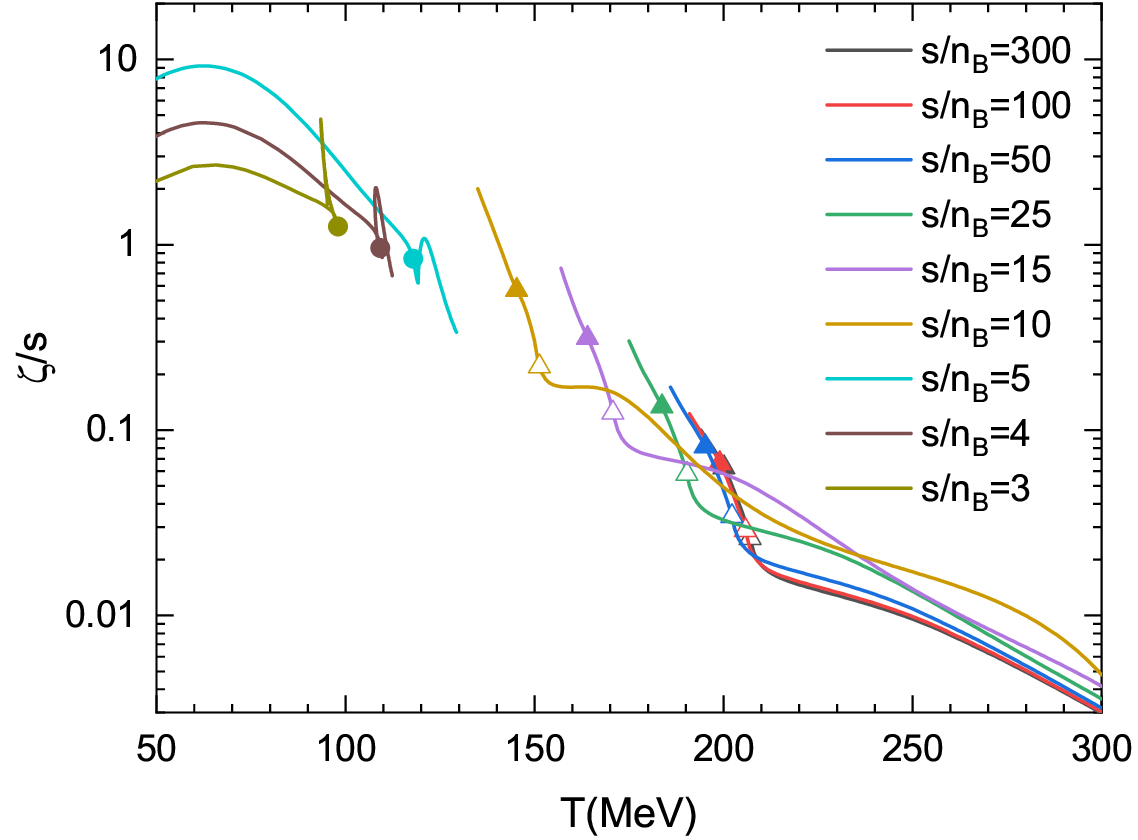}
	\caption{$\zeta/s$ along the isentropic trajectories for different $s/n_B$. The solid triangles, hollow triangles and dots represent  the $\eta/s$ at the points where these isentropic lines intersect the chiral crossover line, pion Mott transition line, the high-density boundary of the first-order transition, respectively. }
	\label{zetassn}
\end{figure}

It is emphasized that under the relaxation time approximation, $\zeta/s$ remains finite at the critical endpoint, with no singular behavior observed. In theory, critical phenomena are generally described by universal dynamical scaling. Near a second-order phase transition~(CEP), transport coefficients such as the bulk and shear viscosities are determined by the dynamical critical scaling specific to a universality class. The QCD critical endpoint falls into model H with the Halperin-Hohenberg  classification~\cite{hohenberg1977theory}, where $\zeta$ is expected to diverge as $\zeta \sim \xi^{z_\zeta}$ with $z_\zeta \approx 3$ and $\xi$ being the correlation length~\cite{moore2008bulk,sasaki2009bulk,sasaki2010transport}. The current work does not include these dynamical critical effects.

Fig.~\ref{zetassn} shows the specific bulk viscosity $\zeta/s$ on the isentropic trajectories. In heavy-ion collision experiments, higher collision energies yield larger values of $s/n_B$, whereas QGP created in relatively lower energies can access the QCD phase diagram at low temperature and high density. Numerical results show that for $s/n_B \geq 10$, the $\zeta/s$ increases monotonically as the temperature decreases, with a more rapid enhancement across the chiral crossover region of $u,d$ quarks. The larger value of $\zeta/s$ in the phase transition region implies greater energy dissipation during volume expansion.
This trend indicates the growing significance of bulk viscosity in hydrodynamic simulations of quark matter generated in lower-energy collisions.  For the cases of $s/n_B = 5$, 4, and 3, where the evolution traverses the spinodal region of the first-order phase transition, the behavior of $\zeta/s$ are much sensitive to the phase structure. One can refer to Ref.~\cite{2024He} for the relationship between these isentropic trajectories and QCD phase structure within the PNJL model.  It can be predicted that the larger  $\zeta/s$ in lower-energy heavy-ion collisions will suppress the transverse momentum of final-state particles in experiments~\cite{2019Bernhard,Denicol09,karsch2008universal,Aboona2025}.

\subsection{Ratio of bulk viscosity to shear viscosity}

The ratio of bulk viscosity to shear viscosity ($\zeta/\eta$) constitutes a  dimensionless parameter that reveals fundamental aspects of the dissipation behavior of QGP. 
The shear viscosity primarily originates from momentum transfer between adjacent flow layers and characterizes the system's resistance to shear deformation. Even in a perfectly scale-invariant system, such as a non-interacting ideal gas of massless particles, $\eta$ remains finite. The change of  $\eta$ with thermal parameters reflects the strength of dynamical interactions and being sensitive to phase transitions. In contrast, the bulk viscosity $\zeta$ quantifies the resistance to volume changes during expansion or compression. As noted previously, a strict scale-invariant system must exhibit $\zeta = 0$. The behavior of $\zeta$ in quark matter thus reflects the degree of scale symmetry breaking induced by strong interactions.
The ratio $\zeta/\eta$ serves as a key measure of the relative importance of these two dissipation mechanisms under different physical conditions.

\begin{figure}[htbp]
	\centering
	\includegraphics[scale=0.4]{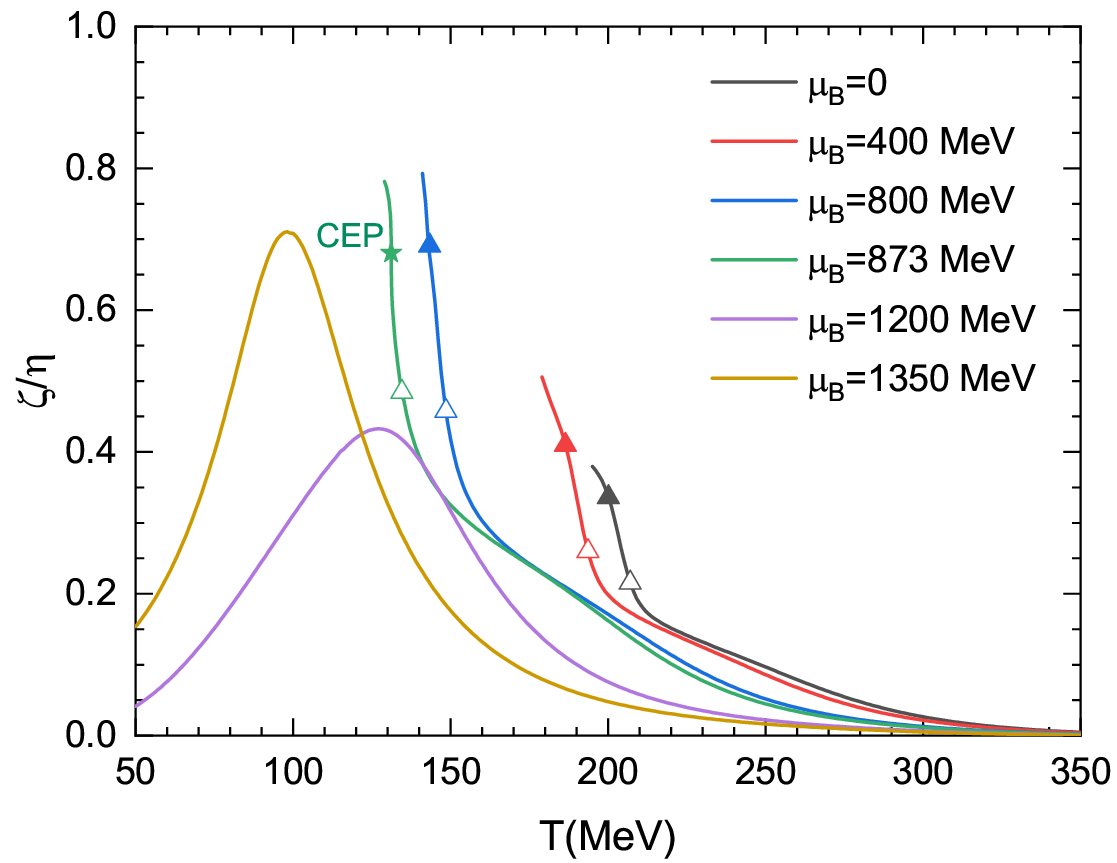}
	\caption{$\zeta/\eta$ as a function of temperature $T$ for $\mu_{B} = 0$, 400, 800, 873, 1200, 1350\,$\mathrm{MeV}$. The star, solid triangles and hollow triangles represent  the $\zeta/s$ at the points where these $\mu_B(T)$ lines intersect the CEP, chiral crossover  line and pion Mott transition line, respectively.}
	\label{zetaetat}
\end{figure}

\begin{figure}[htbp]
	\centering
	\includegraphics[scale=0.4]{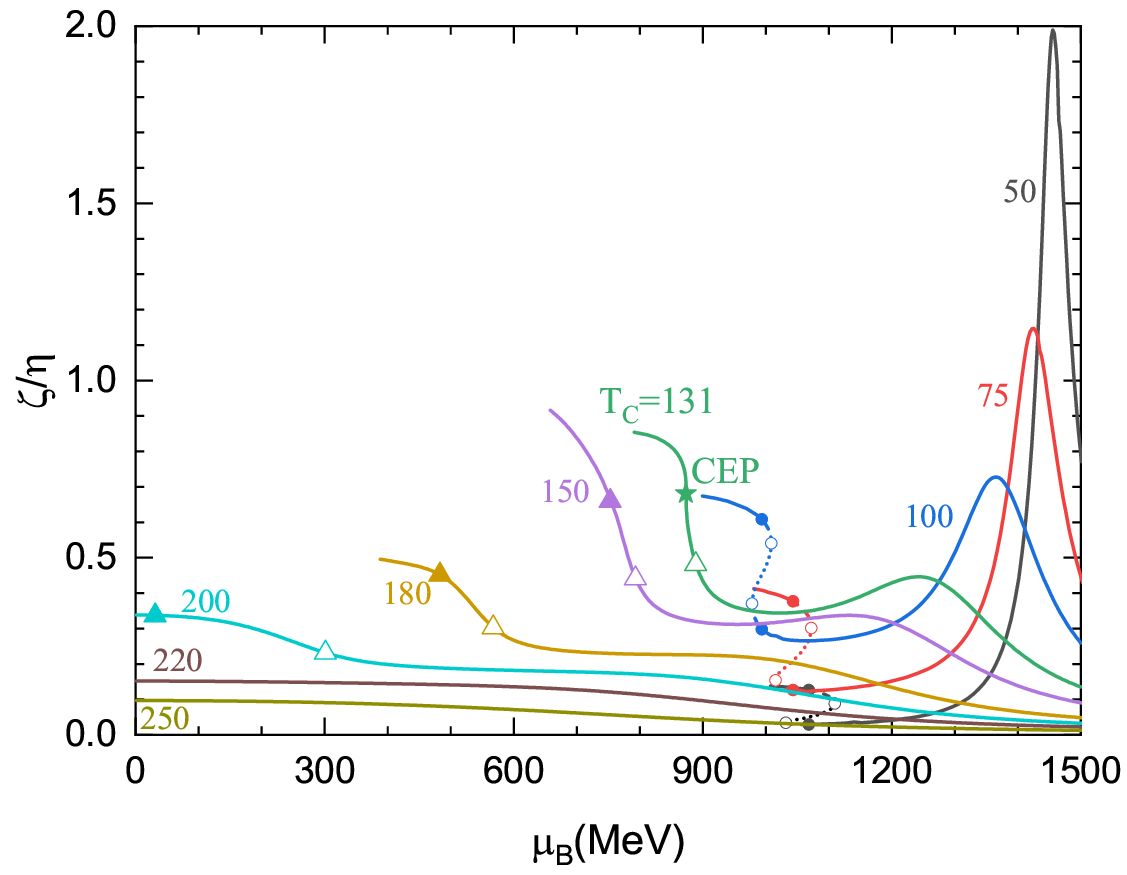}
	\caption{$\zeta/\eta$ as a function of $\mu_{B}$ for different temperatures. The star, solid and hollow triangles represent  the $\zeta/\eta$ at the points where these $T(\mu_B)$ lines intersect the CEP, chiral crossover line and pion Mott transition line, respectively.
	For $T=50, 75, 100\,$MeV, the two dots (cycles) on each isotherm correspond to respectively the high-density chirally restored phase and low-density chirally broken phase on the boundaries (spinodal line) of the first-order phase transition. }
	\label{zetaetamu}
\end{figure}

Fig.~\ref{zetaetat} and Fig.~\ref{zetaetamu} show the ratio $\zeta/\eta$ as functions of temperature and chemical potentail, respectively. One can observe that  $\zeta/\eta$ is small at high temperature and approaches zero in the high-temperature limit. This is because that  the QGP described in the PNJL model restores the approximate scale invariance at high temperatures with the decrease of dynamical quark masses~(near restoration of chiral symmetry). Furthermore,
as shown in Fig.~\ref{zetast}, $\zeta/s$ approaches zero at high temperature, while the ratio $\eta/s$ exhibits a slight increase (Fig. 6 in Ref.~\cite{2024He}).
These features indicate that the energy dissipation from bulk viscous effects is small at extremely high temperature, consistent with the result of conformal invariance.

\begin{figure*}[htbp]
	\centering
	\includegraphics[scale=0.45]{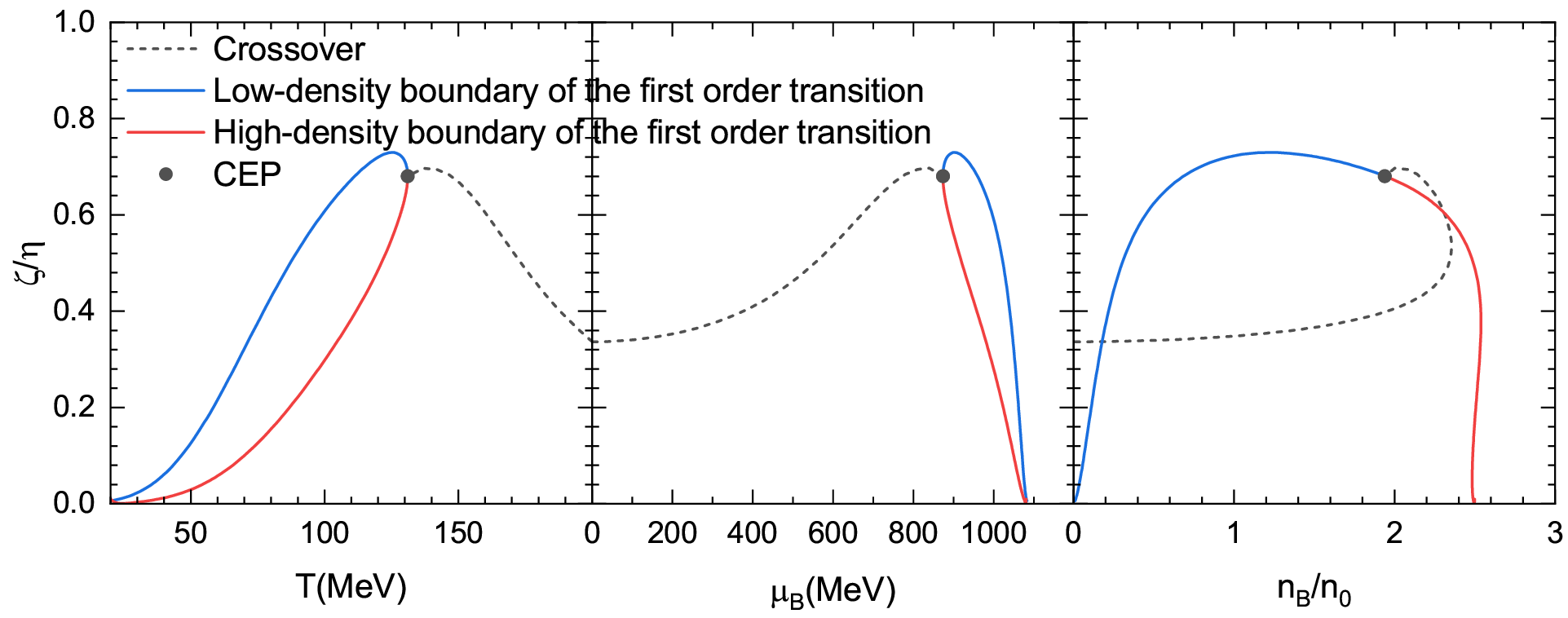}
	\caption{$\zeta/\eta$ on the chiral phase boundaries as functions of temperature, baryon chemical potential and net baryon number density, respectively.}
	\label{zetaetapt}
\end{figure*}

Fig.~\ref{zetast} and Fig.~6 in Ref.~\cite{2024He} also display that both $\zeta/s$ and $\eta/s$ increase significantly near the chiral crossover transformation (marked with solid triangles), which means that the system in this region has  greater energy dissipation from both shear deformation and volume expansion.  Compared to the chirally restored high-temperature regime, the increasing  $\zeta/\eta$ in the chiral crossover region means that the energy dissipation rate due to bulk viscous pressure grows more quickly than that from shear viscosity.

As observed in Fig.~\ref{zetaetat} and Fig.~\ref{zetaetamu}, for the high-density QGP at $\mu_B = 1200$ MeV, the ratio $\zeta/\eta$ increases first and then decreases with the drop in temperature.  This is because that $\eta/s$ grows faster than $\zeta/s$ at low temperature. This feature resembles that of ordinary fluids whose shear viscosity is strongly enhanced at low temperature, making the system more resistant to laminar flow. These results indicate that in the low-temperature and high-density region for $\mu_B = 1200$ MeV, energy dissipation from shear viscosity dominates. The same behavior exists at higher baryon chemical potentials, including the emergence of region where $\zeta/\eta > 1$. The peak shape  in Fig.~\ref{zetaetat} and Fig.~\ref{zetaetamu} are attributed to the chiral crossover transformation  of strange quark. Furthermore, in the regime of chiral symmetry restoration of strange quark at extreme high chemical potentials, $\zeta/\eta$ decreases again.

Fig.~\ref{zetaetapt} presents the value of $\zeta/\eta$ on the chiral phase boundaries. 
Along the chiral crossover line, $\zeta/\eta$ generally increases with decreasing temperature but slightly declines in the vicinity of CEP. For the first-order phase transition, the value of $\zeta/\eta$ on the high-density phase boundary is smaller than that on the low-density boundary.  Near the critical region, the behavior of $\zeta/\eta$ is sensitive to the thermal parameters of temperature, baryon chenical potential and net baryon density.

Notably, $\zeta/\eta$ approaches zero at extreme low temperatures on both sides of the first-order transitions.
In the high-density phase with restored chiral symmetry for $u$ and $d$ quarks, the system is dominated by a highly degenerate Fermi sea. This degeneracy   at very low temperatures strongly suppresses interparticle scattering phase space due to the Pauli blocking effect, leading to a significant enhancement of the shear viscosity $\eta$. Meanwhile, the system becomes ``rigid'' under high pressure and thus approaches scale invariance, which suppresses the bulk viscosity $\zeta$.
In contrast, in the low-density chirally broken phase at low temperatures, the system behaves as a dilute gas of massive particles. The exponential suppression of the particle number density drastically reduces the efficiency of momentum transport, resulting in relatively large shear viscosity, while bulk viscous dissipation from volume changes becomes negligible as the system asymptotically approaches the vacuum. Thus, $\zeta \ll \eta$ emerges as a natural consequence of the system approaching the vacuum.

\section{Summary }

Based on the kinetic theory with relaxation time approximation, we have investigated the specific bulk viscosity ($\zeta/s$)  and the ratio of bulk viscosity to shear viscosity  ($\zeta/\eta$) in quark matter at finite temperature and chemical potential. The in-medium dynamical quark masses  derived in the 2+1 flavor PNJL model are taken to calculate the  collision cross sections under various thermal parameters.

We explored the behaviors of $\zeta / s$  and $\zeta / \eta$ across the QCD phase diagram. The investigation indicates that the values of $\zeta/s$ and  $\zeta/\eta$  have a profound connection with the  QCD phase structure.
Both the $\zeta / s$ and $\zeta / \eta$ are small at high temperatures, consistent with the nature of a conformal theory, whereas larger $\zeta/s$ and $\zeta/\eta$ are derived near the chiral phase transition.
Along the chiral crossover line, $\zeta / s$ and $\zeta / \eta$ generally increase with decreasing temperature, though $\zeta / \eta$ exhibits a slight decline near the CEP. On both the low-density and high-density boundaries of the first-order transition, $\zeta / s$ shows a non-monotonic variation due to  the rates of change for $\zeta$ and $s$ undergoing an inflection with the decrease of temperature.

Furthermore, on the phase boundary, $\zeta / \eta$ attains larger values in the critical region but approaches zero at extremely low temperatures. Notably, beyond the phase boundary, an additional peak structure is observed in both $\zeta / s$ and $\zeta / \eta$, with magnitudes even larger than those on the chiral phase transition boundary of $u,d$ quarks. Our analysis indicates that this prominent peak shape originates from the chiral crossover transformation of strange quark.

It should be also noted that the temperature and chemical potential dependence of $\zeta/s$ and $\zeta/\eta$ obtained in this study are model-dependent in their specific values, nevertheless we expect these results to provide qualitative insight into the bulk viscous behavior of the QCD medium. Furthermore, including additional degrees of freedom of hadrons may improve the result near the chiral crossover. For transport coefficients in the critical area, a proper treatment requires further incorporation of dynamical critical scaling governed by the universality class of the CEP.

\begin{acknowledgments}
	This work is supported by the National Natural Science Foundation of China under
	Grant No. 12475145  and Natural Science Basic Research Plan in Shaanxi Province
	of China (Program No. 2024JC-YBMS-018).
\end{acknowledgments}

\section*{DATA AVAILABILITY}

The data that support the findings of this article are not publicly available. The data are available from the authors upon reasonable request.

\bibliography{ref.bib}
\end{document}